\documentclass[11pt,twoside]{article}  
\usepackage{asp2006}
\usepackage{adassconf}

\usepackage{multirow}

\begin{document}   

%
%

\paperID{P05}

%

\title{Advanced Architectures for Astrophysical Supercomputing}

%
%
%
%
%

\markboth{Barsdell, Barnes and Fluke}{Advanced Architectures for Astrophysical
Supercomputing}

%
%
%
%

\author{Benjamin R. Barsdell, David G. Barnes, Christopher J. Fluke}
\affil{Centre for Astrophysics \& Supercomputing, Swinburne University
of Technology, Hawthorn, Victoria, Australia}

%

\contact{Ben Barsdell}
\email{bbarsdel@astro.swin.edu.au}

%
%
%

\paindex{Barsdell, B.~R.}
\aindex{Barnes, D.~G.}     
\aindex{Fluke, C.~J.}

%



\begin{abstract}          
Astronomers have come to rely on the increasing performance of computers to
reduce, analyze, simulate and visualize their data. In this environment,
faster computation can mean more science outcomes or the opening up of new
parameter spaces for investigation. If we are to avoid major issues when
implementing codes on advanced architectures, it is important that we have a
solid understanding of our algorithms. A recent addition to the
high-performance computing scene that highlights this point is the graphics
processing unit (GPU). The hardware originally designed for speeding-up
graphics rendering in video games is now achieving speed-ups of $O(100\times)$
in general-purpose computation -- performance that cannot be ignored. We are
using a generalized approach, based on the analysis of astronomy algorithms,
to identify the optimal problem-types and techniques for taking advantage of
both current GPU hardware and future developments in computing
architectures.
\end{abstract}

%
%

\section{Introduction}
Modern astronomy has come to rely heavily on high-performance
computing (HPC). However, all research areas are facing significant
challenges as data volumes approach petabyte levels. For instance, the
Australian Square Kilometre Array Pathfinder project will produce
data at a rate that makes storage in raw form impractical,
necessitating on-the-fly reduction and analysis to produce 4GB/s of
products. On the modeling front, there is an ongoing desire for larger
and more-detailed simulations 
(e.g., the Aquarius simulation by Springel et al. 2008).

The HPC scene has recently witnessed the bold introduction of the graphics
processing unit (GPU)
as a viable and powerful general-purpose co-processor to CPUs. GPUs
were developed to off-load the computations involved in rendering 3D graphics
from the CPU, primarily to benefit video-games. Their
continued development has been driven by the \$60 billion/year
video-game industry. The result of this development can be seen in Figure 1,
which plots the clock-rate of a number of CPUs and GPUs against their
core-count. GPUs appear toward the top of the plot, exhibiting very high
core-counts and performance. 
Since 2005 (an area on the plot we refer to as the ``multi-core
corner''), clock-rates in CPUs have plateaued, and manufacturers 
have instead turned to increasing the number of cores per chip\footnote{This
  has to do with the difficulty in dissipating the heat 
  produced at higher clock-rates.}. One might therefore consider that CPUs are
now becoming progressively more GPU-like, and that current GPUs provide a
picture of future commodity computing architectures.


\begin{figure}
\centering
\epsscale{0.5}
\plotone{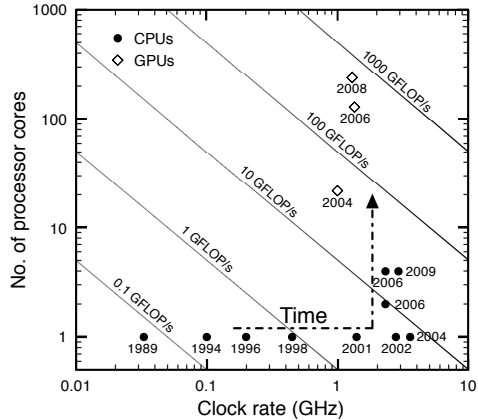}
\vspace{-20pt}
\caption{Clock-rate versus core-count phase space of Moore's Law. Black dots
  represent central processing units (CPUs); white diamonds represent graphics
  processing units (GPUs).
Diagonal lines are contours of equal processing power.
}
\label{fig:MulticoreMooresLaw}
\end{figure}
While CPUs are becoming more GPU-like, the reverse can also be said,
with GPUs offering increasingly
flexible computing platforms. This flexibility, combined with the
availability of general-purpose programming tools, has opened up
GPU computation to a wide range of non-graphics-related tasks, notably
in the area of HPC.

Both the immediate performance boost provided by GPUs and the expected
future of CPU computing provide strong motivation for a thorough
analysis of the performance and scalability of our astrophysics
algorithms in advanced parallel processing environments. If harnessed
correctly, the power of 
massively-parallel architectures like GPUs could
lead to significant speed-ups in computational astronomy and
ultimately to new science outcomes.

A number of astronomy algorithms have been implemented on GPUs
to date, including direct N-body simulations (e.g., Belleman, B\'edorf \&
Portegies Zwart 2008), 
the solution of Kepler's equation (Ford 2008),
radio astronomy
correlation (e.g., Harris, Haines \& Staveley-Smith 2008),
phase-space studies of binary black hole inspirals (Herrmann et al. 2009)
and gravitational lensing ray-shooting (Thompson et al. 2010).
These projects have all reported speed-ups of O(100) over CPU
 codes\footnote{It is understood that these speed-up measurements 
 are unlikely to have been obtained in a consistent manner; we merely
 emphasize their order of magnitude.}.
However, these algorithms are for the most part ``embarrassingly
parallel'' ``low-hanging fruits'', meaning that they can be run on a
parallel processing architecture with little or no overhead. This
makes them obvious candidates for efficient GPU implementation. The
question that remains is: exactly which classes of astronomy
algorithms are likely to obtain significant speed-ups by executing on
advanced, massively-parallel, architectures?

\section{Our Approach}
We propose a generalized approach based around two key ideas:
\begin{enumerate}
\item Developing and applying an algorithm analysis methodology relevant to
  new hardware architectures; and
\item Building and using a taxonomy of astronomy algorithms.
\end{enumerate}
We believe that such an approach will minimise the effort required to
turn the ``multi-core corner'' for computational astronomy and ensure
that the solutions found will continue to scale with future advances
in technology.

Here we briefly outline a number of ``rules of thumb'' that may be applied
when analyzing astronomy algorithms with respect to their potential on
GPUs. Further details will be presented in Barsdell, Fluke \& Barnes (in prep.).

\textbf{Massive Parallelism:} Given the large number of processing cores
available in GPUs, it is critical 
that an algorithm be divisible into many \textit{fine-grained} parallel
elements in order to fully utilize the hardware (e.g., an NVIDIA GT200-class
GPU  may be under-utilized with $<O(10^4)$ threads).
Partitioning data, rather than tasks, between parallel threads
generally offers a large and scalable quantity of parallelism. This is
referred to as the ``data-parallel'' approach.


\textbf{Memory Access Locality and Patterns:} GPU architectures contain very
high bandwidth main memory, $O(100$GB/s$)$, 
which is necessary to ``feed'' the large number of parallel processing
units. However, high latency (i.e., memory transfer startup) costs mean that
performance depends strongly on the \textit{pattern} in which memory is
accessed.
In general, maintaining ``locality of reference'' (i.e., neighboring threads
accessing similar locations in memory) is vital to achieving good
performance.


\textbf{Branching:} GPUs 
contain ``single instruction multiple data'' (SIMD) hardware.
This means that neighboring threads 
that wish to execute different instructions must
wait for each other to complete the divergent code section before execution
can continue in parallel.
For this reason, sections of GPU code that are conditionally executed by only
a subset of threads should be minimized.

\textbf{Arithmetic Intensity:} Executing arithmetic instructions is generally
much faster than accessing 
memory on GPU hardware and thus increasing the number of arithmetic
operations per memory access can help to hide memory latencies. This is not
always possible, and some algorithms will remain bandwidth-limited rather than
instruction-limited. However, this is a case where a more drastic re-think of
a problem may be required for an efficient solution. For example, the
optimal order of a numerical expansion may be different on a GPU architecture
than on a CPU architecture.

\textbf{Host--Device Memory Transfers:} GPUs and their host machines
(typically) have distinct memory 
spaces, meaning they must communicate via the PCI-Express bus, which exhibits
relatively low bandwidth (currently $\sim$5GB/s). Transferring data to and from a
GPU device can therefore be a significant performance bottle-neck in some
situations.

\section{Algorithm Classification}
Here we present an initial classification, based on application of the ``rules
of thumb'' and reduction to known GPU-efficient algorithms, of a selection of important
astronomy problems.
\textit{High efficiency}
algorithms correspond to expected $O(100\times)$ speed-ups, while
\textit{moderate efficiency} algorithms are expected to exhibit speed-ups of
$O(10\times)$ over traditional CPU implementations:
\newline

\renewcommand\arraystretch{1.2}

\begin{centering}
\footnotesize
\begin{tabular}{lll}

\hline
\hline
Field & High efficiency & Moderate efficiency \\
\hline
\multirow{5}{*}{Simulation} & $\bullet$ Direct N-body & $\bullet$ Tree-code N-body and SPH \\
 & $\bullet$ Fixed-resolution mesh simulations & $\bullet$ Halo-finding \\
 & $\bullet$ Semi-analytic modelling & $\bullet$ Adaptive mesh refinement \\
 & $\bullet$ Gravitational lensing ray-shooting & \\
 & $\bullet$ Other Monte-Carlo methods & \\
\hline
\multirow{5}{*}{Data reduction} & $\bullet$ Radio-telescope signal correlation & $\bullet$ Pulsar
  signal processing \\
 & $\bullet$ General image processing & $\bullet$ Stacking/mosaicing \\
 & $\bullet$ Flat-fielding etc. & $\bullet$ CLEAN algorithm \\
 & $\bullet$ Source extraction & $\bullet$ Gridding visibilities and \\
 & $\bullet$ Convolution and deconvolution & single-dish data \\
\hline
\multirow{4}{*}{Data analysis} & $\bullet$ Machine learning & $\bullet$ Selection via
  criteria-matching \\
 & $\bullet$ Fitting/optimisation & \\
 & $\bullet$ Numerical integration & \\
 & $\bullet$ Volume rendering & \\
\hline
\end{tabular}
\end{centering}
\newline

\section{Conclusion}
Modern astronomy relies heavily on HPC, and GPUs can provide both significant
speed-ups over current GPUs and a glimpse of the probable future of commodity
computing architectures. However, their more complex design means algorithms
must be considered carefully if they are to run efficiently on these advanced
architectures. There is therefore strong motivation to thoroughly analyze and
categorize the algorithms of astronomy in order to take full advantage of
current and future advanced computing architectures and maximize our science
outcomes.

Our preliminary analysis of a broad selection of important astronomy problems
leads us to conclude that the data-rich nature of computational astronomy,
combined with the efficiency of data-parallel algorithms on current
GPU hardware, make for a very promising relationship with current and
future massively-parallel architectures. Processors are likely to
become even more flexible in the future, potentially improving the
efficiency of many astronomy algorithms and opening up new avenues to
significant speed-ups.

\end{document}